\documentclass[acmsmall,screen,authorversion]{acmart}
\usepackage{xcolor,colortbl}
\usepackage{graphicx}
\usepackage{color}
\usepackage{pifont}
\usepackage{ifthen}
\usepackage{multirow}
\usepackage{booktabs}
\usepackage{paralist}
\usepackage[inline]{enumitem}
\usepackage[ruled,linesnumbered]{algorithm2e}
\usepackage{tcolorbox}
\usepackage{booktabs}
\usepackage{tabularx}
\newcolumntype{L}{>{\raggedright\arraybackslash}X}%
\newcolumntype{R}{>{\raggedleft\arraybackslash}X}%
\newcolumntype{C}{>{\centering\arraybackslash}X}%
\usepackage{listings}
\usepackage{relsize}
\usepackage{threeparttable}
\usepackage{amsmath}
\usepackage{mathtools}
\usepackage{tikz}
\usepackage{fontawesome}
\usepackage{caption}
\usepackage{subcaption}
\usepackage{titlesec}
\usepackage{xspace}

\usepackage{gnuplottex}

\usetikzlibrary{shapes,positioning}
\makeatletter
\let\old@lstKV@SwitchCases\lstKV@SwitchCases
\def\lstKV@SwitchCases#1#2#3{}
\makeatother
\usepackage{lstlinebgrd}
\makeatletter
\let\lstKV@SwitchCases\old@lstKV@SwitchCases

\lst@Key{numbers}{none}{%
    \def\lst@PlaceNumber{\lst@linebgrd}%
    \lstKV@SwitchCases{#1}%
    {none:\\%
        left:\def\lst@PlaceNumber{\llap{\normalfont
                \lst@numberstyle{\thelstnumber}\kern\lst@numbersep}\lst@linebgrd}\\%
        right:\def\lst@PlaceNumber{\rlap{\normalfont
                \kern\linewidth \kern\lst@numbersep
                \lst@numberstyle{\thelstnumber}}\lst@linebgrd}%
    }{\PackageError{Listings}{Numbers #1 unknown}\@ehc}}
\makeatother

\def\editmode{}

\usepackage{tcolorbox}

\newtcolorbox{custombox}[1]{
	colback=gray!10,
	colframe=gray!70,
	left=1.5mm,
	right=1.5mm,
	top=1.5mm,
	bottom=1.5mm,
	fonttitle=\bfseries,
	arc=0mm,
	leftrule=1mm,
	rightrule=0mm,
	toprule=0mm,
	bottomrule=0mm,
	notitle,
	before=\par\medskip\medskip\noindent,
	before upper={\textbf{#1: } },
}

\ifthenelse{\isundefined{\editmode}}{
\newcommand{\editnote}[3]{%
}

}{
\newcommand{\editnote}[3]{\xspace\colorbox{#1}{\sffamily \smaller \textcolor{white}{~\faCommenting{}~#2~}}\textcolor{#1}{~#3}\xspace}

}

\definecolor{nord0}{HTML}{2E3440}
\definecolor{nord1}{HTML}{3B4252}
\definecolor{nord2}{HTML}{434C5E}
\definecolor{nord3}{HTML}{4C566A}
\definecolor{nord4}{HTML}{D8DEE9}
\definecolor{nord5}{HTML}{E5E9F0}
\definecolor{nord6}{HTML}{ECEFF4}
\definecolor{nord7}{HTML}{8FBCBB}
\definecolor{nord8}{HTML}{88C0D0}
\definecolor{nord9}{HTML}{81A1C1}
\definecolor{nord10}{HTML}{5E81AC}
\definecolor{nord11}{HTML}{BF616A}
\definecolor{nord12}{HTML}{D08770}
\definecolor{nord13}{HTML}{EBCB8B}
\definecolor{nord14}{HTML}{A3BE8C}
\definecolor{nord15}{HTML}{B48EAD}

\definecolor{boxbg}{RGB}{240, 240 ,240}

\newcounter{rq}

\tcbuselibrary{breakable}

\titleformat*{\paragraph}{\bfseries}

\hypersetup{
  pdftitle={The Future of AI-Driven Software Engineering},
  pdfauthor={Valerio Terragni; Annie Vella; Partha Roop; Kelly Blincoe},
  pdfsubject={This paper presents a vision of AI-driven software engineering and discusses key challenges and opportunities for research and practice across the software development lifecycle.},
  pdfkeywords={Software Engineering; Artificial Intelligence; Machine Learning; Large Language Models; APIs; Libraries; Software Testing; Requirements Engineering}
}

\begin{document}

% metadata
% title, conference, authors, keywords, etc.

\title{The Future of AI-Driven Software Engineering}

\setcopyright{acmlicensed}
\acmJournal{TOSEM}
\acmYear{2025} \acmVolume{1} \acmNumber{1} \acmArticle{1} \acmMonth{1}\acmDOI{10.1145/3715003}




\author{Valerio Terragni}
\orcid{0000-0001-5885-9297}
\email{v.terragni@auckland.ac.nz}
\affiliation{%
	\institution{The University of Auckland}
	\city{Auckland}
	\country{New Zealand}
}

\author{Annie Vella}
\orcid{0009-0007-5180-9618}
\email{avel920@aucklanduni.ac.nz}
\affiliation{%
	\institution{The University of Auckland}
	\city{Auckland}
	\country{New Zealand}
}

\author{Partha Roop}
\orcid{0000-0001-9654-5678}
\email{p.roop@auckland.ac.nz}
\affiliation{%
	\institution{The University of Auckland}
	\city{Auckland}
	\country{New Zealand}
}

\author{Kelly Blincoe}
\orcid{0000-0003-4092-9706}
\email{k.blincoe@auckland.ac.nz}
\affiliation{%
	\institution{The University of Auckland}
	\city{Auckland}
	\country{New Zealand}
}

% ACCORDING TO ACM STANDARDS we need to put et al. if the auhtors are more than two
\renewcommand{\shortauthors}{Terragni et al.}

\begin{abstract}
A paradigm shift is underway in Software Engineering, with AI systems such as LLMs playing an increasingly important role in boosting software development productivity. This trend is anticipated to persist. In the next years, we expect a growing symbiotic partnership between human software developers and AI. The Software Engineering research community cannot afford to overlook this trend; we must address the key research challenges posed by the integration of AI into the software development process. In this paper, we present our vision of the future of software development in an AI-driven world and explore the key challenges that our research community should address to realize this vision.
\end{abstract}

\maketitle

\section{Introduction}

In the dawn of computing (1940s), programmers  wrote machine code, consisting of binary instructions to directly program computers' hardware. It was quickly understood that programming \textbf{needed a higher level of abstraction from the hardware}~\cite{barnett1970high}. This allowed programmers to write code that is more readable, understandable, and portable across different hardware. From assembly language (a more human-readable representation of machine code) to scripting languages (e.g., Python and JavaScript), the past 70 years of programming languages and practices have witnessed a continuous pursuit of a higher level of abstraction~\cite{gabbrielli2023programming}. This is to increase developers' efficiency and at the same time cope with the demand for increasingly complex software systems.

\smallskip
While the introduction of high-level programming languages has played a major role in allowing software developers to write concise and expressive code, a paradigm shift occurred in the early 2000s with the widespread use of \textbf{APIs (Application Programming Interfaces) and libraries}. Before that, programmers had to write extensive amounts of code to perform even basic tasks. 
%Each piece of functionality required manual coding, leading to longer development times, increased complexity, and a higher likelihood of introducing faults. 
The shift towards using APIs and libraries had a profound impact on the efficiency and capabilities of software development~\cite{zhong2017empirical,lamothe2021systematic}. Programming can now be informally summarised as \textit{chaining the inputs and outputs of API calls}, allowing an even higher level of abstraction. 

\smallskip
The intuitive, informative, and concise nature of variable and API names is bringing our programs closer to resembling \textbf{human language}. Additionally, the ongoing evolution of higher-level programming languages unmistakably demonstrates a trend towards making language constructs more closely aligned with human speech~\cite{gabbrielli2023programming}. \emph{Can this trend continue and eventually programming will reach the pinnacle of abstraction: natural language?} This is very unlikely. Human speech lacks the basic criteria of programming languages (e.g., lack of ambiguity). However, this does not mean that software engineers can not be aided in writing programs by specifying their intent in natural languages. Developers have been using \textbf{StackOverflow.com~(SO)}, 
to search for solutions of programming tasks using natural language as queries.
Indeed, SO and similar Q\&A websites for developers~\cite{StoreyFOSE2014} have become crucial tools to boost developer productivity~\cite{mao2015survey,philip2012software,sadowski2015developers,sim2011well,stolee,mao2015survey,terragni-issta-2016}. 

%\smallskip
%In contrast to software design, hardware design has immensely benefited from silicon compilation combined with the reuse of pre-designed components, called intellectual property (IP) blocks to form system on chips (SoCs)~\cite{sinha2014correct}, which are now ubiquitous in many computing platforms such as mobile phones and game consoles. Hence, the speed at which new hardware can be designed from IP blocks to create new SoCs, has progressed at a phenomenal pace. This is already going to get a significant boost through AI, where a complex chip can be designed using AI, thus enabling non-hardware experts to be able to design chips. For example, conversational approaches using natural language can be leveraged to design chips using an interactive approach~\cite{blocklove2023chip}. How can this be leveraged by software engineers in a similar manner?

\smallskip
The recent rise of \textbf{Large Language Models (LLMs)}~\cite{zhao2023survey}, especially following the global launch of GPT3.5, GPT4o and more recently, o1, by OpenAI, have brought another revolution of programming, rapidly overshadowing platforms like SO~\cite{da2024chatgpt}. 
While program synthesis from natural language queries has been a subject of research for many years~\cite{gulwani2017program}, the performance of recent LLMs has shown results that were unthinkable just a few years ago~\cite{hou2024systematic,chen2021evaluating,du2024evaluating,fan2023large}. 
Now, developers no longer need to search on SO for code snippets; instead, they can directly ask GPT (or other LLMs), and even have conversational interactions to better understand and improve the generated code. Recently, SO removed statistics on its daily visit counts and officially addressed concerns about declining website traffic in a blog post\footnote{\url{https://stackoverflow.blog/2023/08/08/insights-into-stack-overflows-traffic/}}. The post acknowledges the decline in visits and attributes the trend to the release of GPT-4.
We are witnessing a paradigm shift in software development where software engineers use LLMs and other AI systems to boost their productivity~\cite{rajbhoj2024accelerating,ebert2023generative}. 
We can confidently say that LLMs, alongside high-level programming languages, libraries, and developer Q\&A websites, have become essential tools for modern software development~\cite{ebert2023generative}.

\smallskip
LLMs are here to stay. 
Indeed, their capabilities and performance in a wide range of software engineering tasks are set to improve in the future. This is due to the increasing availability of open-source code for training purposes, alongside the ongoing efforts of the AI community to enhance LLM performance. 
As such, over the next decade, we anticipate that software engineers will continue to use LLMs (or similar AI systems) in software development. 

\smallskip
Our research community must acknowledge and address the opportunities and challenges that arise from the use of AI in software engineering. Concerns persist regarding the quality of AI-generated code~\cite{liu2024your}, with notable issues regarding security and privacy~\cite{yao2024survey}. 

Yet, there are numerous opportunities presented by the versatile capabilities of LLMs, especially when fine-tuned for specific tasks, code bases, or company practices. 
For example, recent research demonstrates that fine-tuned LLMs outperform general-purpose LLMs in code review tasks~\cite{pornprasit2024fine,nashaat2024towards}.
Indeed, software engineering involves much more than writing code. LLMs have proven highly effective in various software engineering tasks beyond code generation~\cite{hou2023large}, including documentation generation~\cite{luo2024repoagent,geng2024large}, testing~\cite{schafer2023adaptive,yuan2023no}, code review~\cite{pornprasit2024fine,nashaat2024towards}, and  program repair~\cite{jin2023inferfix,xia2023automated,xia2022less}.     

Our research community stands at the forefront of this revolution, we need to tempestively address the challenges of the \textbf{symbiotic partnership between human developers and AI}. 

\smallskip
In this paper, we present our vision of the potential future of an AI-driven software engineering, alongside the key research challenges and opportunities associated with the increasing integration of AI into the software engineering process. In particular, we propose the conceptual design of a framework to harness AI capabilities for automating, augmenting, and optimizing various stages of the software development lifecycle, including requirements analysis, design, coding, testing, and maintenance.

\section{A Current Snapshot of AI Tools in Software Development}

This section highlights examples of current tools and platforms that demonstrate how AI (in particular LLMs) are transforming software engineering workflows. While this is not an exhaustive  list, it includes some popular and well-established tools that are widely applicable in development pipelines. Research prototypes are excluded, focusing instead on mature tools ready for real-world use. These examples cover specific applications such as code generation, documentation, and bug fixing, showing how LLMs are becoming integral to modern software engineering practices. We acknowledge that this landscape is shifting very rapidly, and these tools, while the state-of-the-art at the time of writing in December 2024 could be superseded even by the time this article is published.

\smallskip
\textbf{\textsc{GitHub Copilot}\footnote{\url{https://github.com/features/copilot}}} is a  pioneering tool in AI-powered development owned by \textsc{Microsoft}. It is powered by OpenAI’s large language models, and can be integrated with popular IDEs to assist developers by suggesting code completions. It offers context-aware code completions based on the current files and project structure. The 2024 Q2 financial report of  \textsc{Microsoft} reports that \textsc{GitHub Copilot} grew its paid customer base by 30\% quarter-over-quarter to a total of 1.3 million developers and 50,000 organizations\footnote{\url{https://www.microsoft.com/en-us/investor/events/fy-2024/earnings-fy-2024-q2}}. Tools like GitHub Copilot have been found to significantly boost software developer productivity~\cite{ziegler2024measuring}.

\smallskip
\textsc{Amazon} released \textbf{\textsc{CodeWhisperer}}\footnote{\url{https://docs.aws.amazon.com/codewhisperer/}}, its own version of an AI code assistant. Differently from \textsc{GitHub Copilot}, which is general purpose, \textsc{CodeWhisperer} specializes in AWS cloud development, providing tailored guidance and suggestions for AWS-specific coding and infrastructure.

\smallskip
Another recent example is \textbf{\textsc{Windsurf} by \textsc{Codeium}}\footnote{\url{https://codeium.com/windsurf}}, recognized as the first ``agentic IDE'' that seamlessly incorporates AI features into development that go beyond code suggestions. It is a fork of \textsc{Visual Studio Code} that enables developers to:
\begin{inparaenum}[(i)]
    \item Prompt the AI to build an entire application, breaking it down into manageable tasks;
    \item Automatically create files, suggest package installations, and manage dependencies;
    \item Review and refine its own code. %, occasionally correcting errors without user input.
\end{inparaenum}

\smallskip
Similarly, \textbf{\textsc{Cursor}}\footnote{\url{https://www.cursor.com/}}, another fork of \textsc{Visual Studio Code},  automatically fixes generated code and supports app creation by breaking tasks into smaller steps. \textsc{Cursor} offers extensive customization through the \texttt{.cursorrules} file format, enabling developers to adapt AI behavior for specific frameworks and languages.\footnote{\url{https://dotcursorrules.com/}}

\smallskip
LLMs are also reshaping the way documentation is created and optimized. Tools such as \textbf{LLM Text}\footnote{\url{http://llmtext.com/}} help developers extract and summarize relevant context from various sources, including GitHub repositories, npm packages, and YouTube videos. \textbf{UiHub}\footnote{\url{https://uithub.com/}} is another tool that leverages LLMs for software engineering. Its  enables large-scale repository analysis, allowing developers to build advanced development tools and applications.

\smallskip
These concrete examples showcase how current LLMs are effectively being integrated into software engineering, offering developers  powerful capabilities to boost productivity, streamline workflows, and improve code quality. The few examples we discussed unmistakenly show that the evolution of software development is increasingly pointing toward \textbf{autonomous multi-agent systems}, where AI systems independently make decisions during software development~\cite{he2024llm}.

\smallskip
This evolution inspires, motivates, and informs our framework for the future of software engineering, providing evidence that we are heading in that direction. As we embrace this trajectory, it becomes crucial to address the challenges ahead and necessitates rethinking how to better accommodate these intelligent AI-powered agent collaborators.

\section{AI-Driven Software Engineering}

\begin{figure*}[t]
    \centering
    \includegraphics[width=1\linewidth]{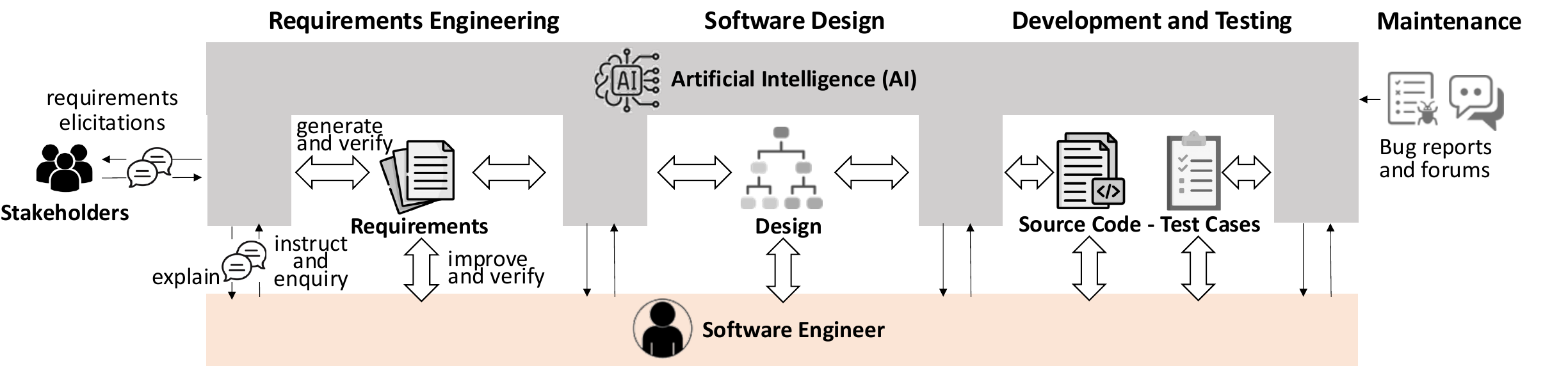}
    \vspace{-3mm}
    \caption{Logical architecture of the envisioned future symbiosis of Software Engineers and AI}
    \label{fig:overview}
\end{figure*}

Figure~\ref{fig:overview} overviews our envisioned \textbf{AI-driven software development framework}. While certain aspects of this framework may appear overly optimistic about the capabilities of future AI systems, it presents an interesting thought process for understanding the potential symbiotic synergy between AI and software developers. Moreover, it sheds light on the research challenges that our community must address to realize this vision someday. Indeed, such a vision is not completely unrealistic. We know that current AI systems can accomplish most of the specified tasks, albeit with limited quality~\cite{huang2024generative,
luo2024repoagent,geng2024large,schafer2023adaptive,yuan2023no,jin2023inferfix,xia2023automated}.

\smallskip
The framework touches all main phases of the \textbf{Software Development Life Cycle}: Requirement Engineering, Software Design, Implementation, Testing, and Maintenance. Note that we are not assuming a waterfall model, the cycles may overlap, especially in agile development methodologies where development cycles are shorter and more flexible. 

\smallskip
\textbf{The main Actors} in our framework are \textbf{software engineers} (e.g., developers, architects, and testers) and a generic AI system (e.g., an LLM). It is important to mention that we believe we are still very far from completely replacing software engineers with prompt engineers. Capable software engineers (with prompt engineering training) will remain indispensable for understanding, reviewing, improving, combining, validating, and maintaining the source code generated by AI. In the short and medium term future, AI is merely a tool to enhance software developers' productivity. While it can automate certain tasks, we assume the presence of humans in the loop.  

\smallskip
Stakeholders (e.g., end users, product owners) are also involved in the framework during the requirement engineering process. AI, equipped with chatbot capabilities, can initiate conversations with stakeholders to elicit, analyse, specify, and
validate requirements.

\smallskip
With our proposed framework, software engineers can either directly create or update the artifacts (i.e., requirements, design, production and test code) or instruct the AI (e.g., through prompt engineering) on how to do that. We envision a \textbf{bi-directional communication} between humans and AI, where humans can ask questions or provide instructions, and the AI can notify the software engineers of any detected issues or opportunities for improvement. Software engineers will communicate with AI through \textbf{conversational interactions} facilitated by the conversational capabilities of LLMs. This interface empowers engineers to seek clarifications and explanations about the artifacts as well as the AI system's output~\cite{liang2024alarge}.

\smallskip
Another important clarification is that, for simplicity, Figure~\ref{fig:overview} represents a single AI system. In reality, different tasks would be handled by specialized AI agents~\cite{xi2023rise}, each fine-tuned for its particular function. These subsystems may either operate autonomously or require direct input from software engineers, depending on the nature of the task.

\smallskip
In particular, the AI agents must effectively communicate with each other and with various software analysis tools responsible for gathering information on the software artifacts in development. As the number of available AI systems continues to grow, to prevent information overload, humans will interact with a single \textbf{unified interface}. Similar to mediator bots~\cite{ribeiro2022together}, an \textbf{orchestrator of AIs} can efficiently manage all interactions with the AI agents behind the scenes.
We envision that the AI's orchestrator will constantly monitor changes in the artifacts (after every update from engineers or from the agents) and invoke the dedicated AI agent to check for consistency and integrity of the artifacts.

\smallskip
From a technical perspective, our framework is a multi-AI agent system~\cite{xi2023rise}. AI agents are artificial entities that can autonomously perceive and act within their environment to achieve specific goals~\cite{xi2023rise}. These agents typically consist of four key components: planning, memory, perception, and action~\cite{xi2023rise}. The planning and memory components serve as the brain controlled by the AI (e.g., LLM), while perception and action enable the agent to interact with its environment and perform specific tasks. While a single-agent system is specialized for solving individual tasks, collaboration between multiple agents (i.e., multi-agent systems) enables the handling of more complex tasks. In the context of software engineering, multi-agent systems can support end-to-end software development processes~\cite{liu2024large,he2024llm}. We envision that the AI orchestrator will coordinate the perception, and action of each agent. We refer readers to the recent survey on Large Language Model-Based Agents for software engineering by Liu et al.~\cite{liu2024large}, which provides a comprehensive review of LLM-based agents in software engineering.

\smallskip
Achieving such a symbiotic partnership between human developers and AI presents several challenges. First, we must not underestimate the difficulties that the orchestrator will face in adapting LLM-based agents to fit existing software engineering tools, processes, and practices. It remains unclear whether entirely new processes and workflows are needed or if LLM-based agents can integrate seamlessly into current ones. Additionally, software developers need training to effectively use LLMs, including recognizing when they are hallucinating. While the framework will automate most interactions with the AI, we believe communication between developers and AI will remain essential—especially when humans seek explanations for the AI's outputs or wish to provide creative insights to improve the software artifacts. To achieve this, developers must be proficient in prompt engineering to communicate effectively with LLMs and guide them toward desired outputs. This is a completely new skill for developers and should be incorporated into existing software engineering education curricula. Furthermore, while AI will undoubtedly be valuable in automating repetitive tasks, human creativity will continue to play a crucial role. Ensuring that creativity remains a dominant force in the workflow presents a significant challenge. Recognizing when AI requires human input versus when it can act independently will be critical.

\begin{custombox}{AI Multi-Agent System}
The primary research challenge in integrating AI into the software development process will be orchestrating the various AI subsystems that focus on specific development tasks and seamlessly integrating them using a single human-AI interface.
\end{custombox}

\subsection{Requirements Engineering}

Gathering requirements accurately is crucial for the success of any software project~\cite{young2004requirements}. Unclear or incomplete requirements can lead to misunderstandings, delays, and ultimately project failures. When requirements are not effectively gathered, developers will build features that do not align with stakeholder expectations and needs, resulting in wasted time and resources. However, this process is often challenging, as \textbf{understanding stakeholder needs} is a complex activity. For example, ambiguities in natural language, stakeholders not always knowing what they truly need or want, and evolving expectations contribute to this complexity. Moreover, translating complex functional or non-functional requirements into technical specifications can be difficult. This makes the requirements elicitation phase complex, time-consuming, and prone to errors.

\smallskip
Recent studies show that AI, particularly LLMs, can assist in requirements engineering activities~\cite{marques2024using,lubos2024leveraging,arora2024advancing}. These models are capable of analyzing, organizing, and summarizing large amounts of data, and, as a result, they can play a crucial role in the preliminary phase of requirements elicitation~\cite{marques2024using}. Stakeholders can provide documentation in any form, and LLMs can summarize large documents or translate them into formal requirement specifications. In particular, we envision AI supporting three key requirements engineering activities: elicitation, validation, and summarization.

\smallskip
\textbf{AI-Assisted Requirements Elicitation:} 
AI-driven agents, such as chatbots powered by LLMs, could assist by engaging stakeholders in human-like conversations to elicit requirements. These agents are capable of generating clarifying questions and suggestions to help stakeholders articulate their needs more clearly. Moreover, they can propose relevant examples or scenarios to facilitate discussions and clarify ambiguities. For example, AI agents could produce mockups of interfaces or rapid prototypes to confirm understanding of user needs. This is particularly useful when stakeholders describe their envisioned solutions to a problem rather than the underlying problem. AI systems must ensure that stakeholders' proposed solutions do not limit innovative designs. 

\smallskip
Dependency parsing can further enhance the elicitation process by structuring stakeholder input, identifying ambiguities, and generating follow-up questions to refine unclear requirements. It can also classify functional and non-functional requirements, ensuring that captured requirements are both structured and actionable~\cite{dalpiaz2019requirements}.

\smallskip
Given the complexity and importance of requirements elicitation, we believe software engineers will remain in the loop. They should oversee these AI-stakeholder interactions in real time or review them afterward, refining and validating the gathered requirements, and intervening when necessary to improve the elicitation process. While AI can facilitate communication, engineers are essential for interpreting non-verbal cues and ensuring stakeholders fully understand the discussion~\cite{wheatcraft2018communicating}. However, future advancements in \textbf{multimodal models} could allow AI systems to better interpret non-verbal cues such as facial expressions and gestures, enhancing their ability to fully capture stakeholder intent. This would reduce some of the reliance on engineers for this aspect, although human oversight would likely remain crucial for more complex, high-stakes discussions. This further emphasizes the need for an engineer’s involvement in the elicitation phase, especially in cases where nuanced human communication is critical. 

\smallskip
\textbf{AI-Assisted Requirements Validation:} 
While AI can contribute significantly during the elicitation phase, its true value lies in the validation of requirements. AI can continuously and automatically analyze requirements to detect inconsistencies, contradictions, and ambiguities. Using Named Entity Recognition (NER), AI can assist in refining these specifications by identifying vague references or conflicts in Software Requirements Specifications (SRS), ensuring consistency and reducing ambiguity~\cite{malik2024supervised}. More importantly, AI can regularly monitor evolving software artifacts (e.g., source code, documentation, tests) to ensure alignment with the requirement and suggest corrective actions when needed.
Additionally, as AI systems increasingly participate in validation, it’s important to ensure that their outputs are ethically sound and unbiased. Regular audits and transparency in how AI reaches decisions will help prevent biased recommendations from influencing projects, ensuring that validation processes remain fair and accurate.
By maintaining a link between the requirements and software artifacts, AI could also predict the potential impact of changes in requirements by analyzing dependencies across the project, helping to assess risks of changing the requirements.

\smallskip
\textbf{Requirement Summarisation and Refinement:} LLMs can assist by summarizing long or complex requirements, allowing engineers and stakeholders to focus on the most important details. By processing large volumes of text, LLMs can extract key points and streamline the language, making it easier to identify priorities. A challenge here will be to ensure prioritisation is fair and considers needs of minority groups (e.g., accessibility concerns). The natural language processing capabilities of AI can also help improve clarity by simplifying complex phrasing and ensuring consistent terminology is used across all documents and discussions, reducing the risk of misunderstandings between teams.

\smallskip

\begin{custombox}{Requirement Engineering}
The main research challenge will be to enable AI agents that can understand user needs to effectively validate the requirements ensuring accuracy and completeness. 
\end{custombox}

\subsection{Software Design}

The integration of AI in software engineering holds immense potential in assisting with the design phase. Building on well-defined requirements, AI can assist in proposing initial design suggestions to align functional and non-functional needs. These suggestions can serve as starting points for further refinement and validation by the engineers and stakeholders. We believe human involvement will remain essential in this step, as the design phase should not—and does not need to—be fully automated. Design artifacts are a vital communication tool between engineers and stakeholders. Additionally, AI can aid in design validation by generating visual artifacts, such as UML diagrams~\cite{abdelnabi2020generating} or C4 models, helping engineers explore different solutions and ensure they meet the project’s requirements. These design artifacts, ranging from prototypes to models and diagrams, offer a more structured and less ambiguous way to communicate design intentions compared to using natural language alone (requirements). We envision several key ways in which AI can enhance the design process and improve communication between engineers and stakeholders.

\smallskip
\textbf{Prototypes sketching:} Requirements written in natural language, while essential for early-stage communication, often suffer from ambiguity, especially when describing complex systems. Prototypes and sketches offer a way to mitigate this issue by providing visual and structural representations of software design. AI-driven tools can automatically generate such artifacts based on requirements. These prototypes could be as simple as sketches, serving as a tangible starting point for discussions among engineers and stakeholders. This approach mirrors the use of paper sketches in traditional design processes, making early-stage design clearer and more accessible to non-technical audiences. This allows for early feedback and adjustments, accelerating the design iteration process. By creating these early-stage prototypes, AI not only aids in communication but also helps in validating ideas against requirements and constraints. AI can greatly improve this process by quickly and automatically generating mock UIs and tangible artifacts.

\smallskip
\textbf{Multi-Level Design Artifacts for Different Stakeholders:} One of the critical challenges in software design is communicating with stakeholders who have varying levels of technical expertise. While developers may require detailed, low-level design documents, project managers and business stakeholders often benefit from high-level, conceptual overviews. AI can facilitate the creation of multi-level design artifacts that take into account these different audiences. For instance, AI could generate a high-level system overview using the C4 model\footnote{C4 Model: https://c4model.com/}, which provides four layers of abstraction for system design: context, containers, components, and code. At the same time, the AI system could automatically produce more detailed, low-level diagrams for developers (e.g., UML class diagrams, UML sequence diagrams, state machine diagrams, component and deployment diagrams). By maintaining consistency across these different levels, AI will help ensure that all stakeholders, regardless of technical proficiency, have a clear and accurate understanding of the design. Moreover, AI can personalize the level of detail provided based on the user’s role, dynamically adjusting the complexity of design artifacts. This adaptability enhances communication and collaboration between technical and non-technical teams, making software development more inclusive and efficient. We envision fine-tuned LLMs trained on design artifacts, source code, and requirements seamlessly generating and synchronizing these different levels of design artifacts.

\smallskip
\textbf{Maintaining Up-to-Date Design Documentation:} One of the persistent challenges in software development is keeping software  design documentation synchronized with the actual codebase, especially as changes are made during development. AI could automate this process, ensuring that design artifacts reflect the current state of the system. For example, an AI system could be integrated with version control tools to automatically update architecture diagrams when code modifications are committed. This eliminates the need for manual diagram updates, which are often overlooked, leading to outdated and potentially misleading documentation. AI-driven automation of this process enhances the reliability of design documentation and ensures it remains useful throughout the development lifecycle. 

\smallskip
\textbf{Validating Design Solutions:} Beyond creating and maintaining design artifacts, AI can also contribute to the validation of design solutions. AI systems, trained on best practices and past project data, can analyze proposed designs for flaws or inefficiencies. For instance, AI could simulate different architectural choices and highlight potential bottlenecks, performance issues, or security vulnerabilities in early design stages. By using AI to validate designs, software engineers can receive real-time feedback on their decisions, reducing the risk of encountering major issues later in the development process. AI can also help in identifying trade-offs between different design alternatives, such as balancing non-functional requirements, for example balancing performance with maintainability or scalability. The ability to foresee potential problems in design helps developers make more informed, data-driven decisions, leading to more robust and efficient systems. 

\smallskip
\textbf{Explainability and Trust in AI Design Tools:} A critical factor in the successful integration of AI in software design is explainability. While AI systems can propose design suggestions or evaluate trade-offs, these recommendations must be accompanied by clear, interpretable explanations. Explainable AI (XAI)~\cite{dwivedi2023explainable} techniques are essential to build trust among developers, who must understand the reasoning behind AI-generated suggestions to make informed decisions. Research into XAI is particularly important in the context of software design, where trust and transparency are paramount. AI should be able to justify its choices, explain the benefits and drawbacks of each design alternative, and provide insights into the long-term implications of design decisions. This transparency will help developers feel more confident in relying on AI for complex design tasks and foster greater collaboration between humans and machines. Explainable AI is an important and active research topic in the AI community~\cite{xu2019explainable,dwivedi2023explainable}, and more work is needed to leverage explainability techniques in the context of software design. 

\begin{custombox}{Software Design}
An important research challenge will be to understand how software engineers can effectively integrate AI into their design workflows, communicate with them, and interpret their suggestions. In particular, AI must provide explanations for their design suggestions to increase trust and facilitate human understanding.
\end{custombox}

\subsection{Software Development and Testing}

%Considering that in later stages LLM could generate code starting from the requirements. We need an additonal agent that converts the requirements in something that LLM can understand well.

%We don't believe that will be necessary to define a new \textbf{prompt-friendly requirement language}. 

%More research on fine-tuning and prompt engineering is needed to understand what are good prompts to specify requirements and at the same time to generate the corresponding source code.

%We will also need define a new \textbf{prompt-friendly requirement language} that can enhance collaboration between humans and AI systems in transitioning from requirement engineering tasks to development tasks. We call this language "prompt-friendly" in the sense that it should be easily understood by LLMs so that they could generate the associated source code. For example, the language might need to unambiguously separate functional and non-functional requirements to help the LLM generate code. 

We envision that software development and testing will be intertwined, as automated testing should be conducted to verify the correctness of the components generated by AI, as well as their seamless integration into the code base. Given a set of unimplemented requirements, AI will automatically generate and test the production code, after which humans and AI will collaborate to improve and verify it. Indeed, AI is already being used to automatically generate code and its associated tests, for example GitHub's Copilot. However, we envision several ways AI code generation will evolve. 

\smallskip
\textbf{Ensuring high quality code:} First, it is important that AI-generated code is correct (behaves as expected). Code generated by existing LLMs is not always correct~\cite{nguyen2022empirical}. We envision significant improvements in the correctness of AI-generated code, particularly as systems learn from the selections and refinements of generated code made by software engineers. Thus, it is also important that code generated is understandable while humans remain in the loop. There is a risk, as more AI generated code is integrated into software systems and is fed back into training data for future code generation, that understandability will be reduced. 

\smallskip
Other quality attributes like reliability, security, scalability, and performance will also be important as more code is generated by AI. This can be particularly challenging given that there will be large variations in the training data when it comes to these attributes. For example, research has found that AI generated code is often insecure, but can be improved by effective prompt engineering~\cite{goetz2024you}. Thus, more work is needed to ensure these quality attributes are considered in AI code generation models and good prompt engineering will continue to be important. Education in software engineering will need to integrate prompt engineering into the curriculum.

\smallskip
\textbf{AI-Assisted updates:} It is well known that software requirements continuously change. Changes can occur for many reasons, including changing user needs, changing environments, or new regulations. The complexity of evolving software to keep up with these changes was recognised decades ago by Lehman~\cite{lehman1996laws}. Agile processes and methods have been created to enable software teams to better respond to these continuous changes~\cite{williams2003agile}. We envision that, as requirements change, AI can also automatically validate the new requirements (as described in the previous section) and generate changes to the associated code and tests. A challenge here is that requirements might be too high level, and it is difficult to decompose high-level requirements into low-level implementation details. However, we believe that as AI advances, we will move further in this direction. 

\smallskip
By producing corresponding tests, the requirements can be automatically verified to ensure the system fulfills the changed or added requirements. This brings promise of more advanced end-user software engineering~\cite{burnett2004end}, where users of software systems can work together with AI systems to specify changes. This could enable more customised and personalised software systems. This is enabled through the advancements we described in previous sections, particularly the ability for AI to validate requirements and produce prototype sketches. With these, stakeholders and users without technical expertise can continuously refine their requests until the desired prototypes are obtained. Automated code and test generation could then proceed to enable these personalisations. However, this also brings maintenance challenges, which we discuss further in the next section.

\smallskip
\textbf{Sharing of validated AI-generated code:} An important opportunity arises from the potential sharing of low-level implementations generated by AI within the open-source community. Low-level implementations could be generated as stateless and immutable APIs. The advantage is that these APIs undergo human and automated verification and testing, including security checks to mitigate vulnerabilities. This enables reuse in other projects rather than regenerating from scratch. By accessing existing databases of AI-generated APIs, AI systems can explore alternatives before generating new code. This concept parallels the notion of "APIzation" recently explored for Stack Overflow code snippets~\cite{terragni-ase-2021,terragni-issta-2016}. However, caution must be taken to ensure adherence to open-source licensing and governance when reusing or sharing AI-generated APIs.

\smallskip
This can also help with sustainability challenges given the high cost of code generation in regards to energy consumption~\cite{guo2024stop}. We see sustainability of AI systems a key research challenge in the coming years. As AI code generation continues to mature and AI capabilities increase, code should be continually improved to ensure quality is maintained. Research efforts can look into how this can be done while reducing energy consumption. 

\begin{custombox}{Software Development}
%The key research challenge will be to understand how effective prompt engineering can guide code generation, particularly when aiming for seamless integration into the code base while matching the design and technologies. Indeed, requirements might be too high level, and it remains a challenge how to decompose high-level requirements into low-level implementation details. 
Ensuring generated code is correct, understandable, reliable, secure, and scalable, while also considering energy consumption of the AI models, is a key research challenge. Prompt engineering will continue to be important. 
\end{custombox}
\medskip

\textbf{Testing will play a crucial role}, as we need to ensure the correctness of the LLM-generated code and its integration with the codebase. Test cases can, of course, be created by developers, but they can also be generated automatically. The latter type of test cases will be crucial for verifying AI-generated code. 

    Researchers are exploring the use of LLMs to generate test cases~\cite{li2024evaluating, bhatia2024unit, dakhel2024effective, schafer2023empirical, xie2023chatunitest,ouedraogo2024llms,yang2024evaluation,lops2024system}, showing promising results. Therefore, we assume that an LLM-based agent dedicated to test case generation will be included in our framework.

However, we envision that such LLM-based agents will work in combination with automated test generators (e.g., \textsc{Randoop}~\cite{pacheco2007randoop}, \textsc{EvoSuite}~\cite{fraser2011evosuite}, and \textsc{Pynguin}~\cite{lukasczyk2022pynguin}) to improve the quality and fault detection effectiveness of the generated tests. We are already witnessing the first attempt of this combination, yielding promising results~\cite{lemieux2023codamosa}. While LLMs can be somewhat effective in generating test cases~\cite{schafer2023adaptive,yuan2023no}, current LLMs do not guarantee compilable or runnable test cases~\cite{yuan2023no}.

\smallskip
Therefore, an integration with traditional test generators that compile and run test cases is necessary. Additionally, the feedback from compiling and running test cases is known to be extremely useful in improving LLM-generated tests~\cite{schafer2023adaptive,yuan2023no}, or automatically generate test cases in general (e.g., feedback-directed approach~\cite{pacheco2007feedback}). More research is needed to better exploit the synergy and complementarity of LLMs and traditional test case generators~\cite{lemieux2023codamosa}.

\smallskip
Indeed, generating \textbf{effective oracles} that correctly distinguish between correct and incorrect executions is crucial.  We cannot expect humans to write oracles for (many) AI-generated test cases; we need automatically generated oracles. Unit test generators (e.g., \textsc{Randoop}~\cite{pacheco2007randoop} and \textsc{Evosuite}~\cite{fraser2011evosuite}) generate (regression) oracles based on the implemented behavior, not the intended one. They capture the implemented behavior of the program with assertions that predicate on the values returned by method calls and fail if a future version leads to behavioral differences. Thus, they are only useful in a regression testing scenario, and their effectiveness is usually evaluated in such a scenario~\cite{comparison,jahangirova2023sbfttrack}. Regarding AI-generated code, the regression scenario is not useful as we want to expose faults in the current version of AI-generated code. 

Although recent research shows that LLM-based agents for test generation often produce oracles that capture the actual program behavior instead of the expected behavior~\cite{konstantinou2024llms,zhangexploring,binta2024togll}, their overall effectiveness remains limited and still far from completely addressing the oracle problem~\cite{konstantinou2024llms}.

\smallskip
\textbf{Metamorphic Testing (MT)}~\cite{1998-chen-tr} could be the key to address this challenge. MT alleviates the oracle problem by using relations among the expected outputs of related inputs as oracles~\cite{2017-chen-cs}. Research shows that such relations, called Metamorphic Relations~(MRs), exist in virtually any software system~\cite{2016-segura-tse}. 
MT proves highly beneficial when integrated into automated test generation, as a single MR can be applied to all test automatically generated inputs that satisfy the input relation. However, MT's automation and effectiveness depend on the availability of MRs. The automated generation or discovery of MRs presents a challenging and largely understudied problem~\cite{2016-segura-tse,2017-chen-cs,chen2021new,ahlgren2021testing}. Only recently has the research community begun addressing metamorphic relation generation from different angles~\cite{automr,2014-zhang-ase,gassertmrs,BlasiGEPC21,genmorph,mrscout, xu2024mrAdopt}. More research is needed on MR generation~\cite{genmorph,gassertmrs,mrscout}
and oracle/generation improvement~\cite{terragni-fse-2020,terragni-icse-2021,oasis,oasistse,evospex} to facilitate effective testing of AI-generated code. 

Interestingly, recent research has explored using LLMs for metamorphic testing of software systems~\cite{wang2024software,siddiq2024using,alshahwan2024automated}, including automatically generating MRs~\cite{shin2024towards,tsigkanos2023large,luu2023can,srinivas2023potential,xu2024mrAdopt}. These studies show the potential of LLMs to fully automate MT.

\begin{custombox}{Software Testing}
The key research challenge will be to automatically generate test cases with effective oracles to verify AI-generated code.
\end{custombox}

\subsection{Software Maintenance}

We envision an AI-powered maintenance phase that remains constantly active in the background. The AI will monitor a wide range of external information sources about the software product and its ecosystem to proactively gather potential issues or opportunities for improvement.

\smallskip
Indeed, issues or maintenance opportunities are often buried in a \textbf{large amount of sources}, such as bug reports, automated alerts, error logs, discussions on developer forums, and feedback from app stores~\cite{van2021role,tizard2022rqmtsecosystem}. While current AI tools like Snyk's DeepCoode focus on real-time code analysis for security vulnerabilities and code quality, we envision future AI systems extending this capability by analysing and aggregating insights from diverse, unstructured data sources. Such systems will be capable of autonomously extracting relevant insights, identifying potential issues, and proposing appropriate fixes or improvements to the software artifacts.

\smallskip
In particular, there are important ethical considerations when new product improvements and feature requests can be gathered from the crowd. The AI system should not solely focus on the most popular feature requests and issues but also those that are less popular but might target minority and disability groups~\cite{malgaonkar2022prioritizing,eler2019android}. Further, the AI cannot simply add every feature users suggest, some consideration with the product strategy must be considered~\cite{knauss2014openness}. AI systems can be trained on the product strategy documents to ensure new features align with the overall vision for the product. The strategy will need to be continuously kept up-to-date and while AI can likely assist with this, humans must be kept in the loop.  

\smallskip
Another way AI can assist in software maintenance is by keeping track of the context and history of a software project. AI can learn from the project's history to avoid repeating mistakes from the past. Detailed comments on design decisions and bug fixes can enable this, and explainable AI, as discussed above, can automatically create such details in the future. 

\smallskip
Additionally, software exists within an \textbf{ecosystem of external libraries}. The libraries that a project depends on may release new versions to address vulnerabilities or bugs, making it essential to keep dependencies up to date. However, major updates can introduce breaking changes or compatibility issues with the existing codebase. The AI system should automate minor and patch updates but carefully assess major upgrades, which often require more adaptation due to potential incompatibilities. By analyzing the project’s dependency history and identifying necessary code modifications, the AI can ensure that security and performance improvements are applied without disrupting existing functionality. Furthermore, the AI should detect and resolve static~\cite{jayasuriya2023understanding,jayasuriya2024emse} or behavioral~\cite{jayasuriya2024behavioral} breaking changes, reducing manual intervention and supporting developers in maintaining stable software. Future research could build on work in automated program repair~\cite{goues2019automated,liu2021critical} to further enhance these capabilities.

\smallskip
As described in the previous section, advancements in AI code generation will enable more sophisticated end-user programming. Users can specify new features to customise and personalise their software systems, leading to challenges in terms of software maintenance due to the potential proliferation of different software versions. AI should be capable of continuously updating these personalised versions, ensuring that new features and optimisations from the main version remain accessible to users who have made customisations. Additionally, gate-keeping mechanisms will likely need to be developed to prevent personalisations from introducing security vulnerabilities or other issues. 

\smallskip
Another potential challenge is to ensure adherence to software licenses. Currently, AI-generated code does not automatically account for the licenses of the software used in its training data, leading to concerns about potential violations. For example, \textbf{Copyleft licenses} require that any modified or extended version of the code remains under the same license and attributes the original source~\cite{broussard2007copyleft}. Tools like GitHub Copilot have faced criticism for reusing open-source code without appropriate attribution or compliance with these licensing terms~\cite{basanagoudar2023copyright}. Companies like OpenAI, GitHub, and Microsoft argue that using publicly available code to train AI systems falls under fair use. However, this legal position remains contentious, and future regulations may require AI-generated code to explicitly include attribution and license details. Addressing these legal challenges would necessitate advancements in AI code generation systems, ensuring they can identify and apply the correct licenses to any code they produce.

\begin{custombox}{Software Maintenance}
The primary research challenge will be to enable AI to autonomously process and utilize a vast amount of external information effectively to identify potential issues or opportunities for improvement. The AI should achieve this while ensuring fairness in its decision-making process and adherence to strategic direction.
\end{custombox}

\section{Limitations and Risks}

% THESE ARE THE COMMENTS OF THE REVIWER
%limitations and risks of LLMs in SE practices (e.g. ethics., limitations of LLMs,…). Adding this content will address these concerns and provide a more holistic view of LLMs' role in SE.
%specific challenges and opportunities that may arise from this synergy when humans are involved too, including ethical implications, and potential limitations of AI in software engineering tasks. 
%But I suggest the authors to discuss some potential limitations of LLMs and chanllenges in intergrating AI to developers' workflow seamlessly. Additionally, the authors may want to highlight the potential risks of using LLMs in software engineering, e.g., it may memorize many software secrets and expose them to other developers, etc.

The synergistic collaboration between software developers and AI (in particular, LLMs) offers immense benefits. For example, automating repetitive tasks or augmenting human effort can allow developers to focus on higher-level design and problem-solving. However, it also comes with certain limitations and risks. 

\subsection*{LLM-Specific Limitations and Risks}

\medskip
\noindent \textbf{Correctness Issues:} LLMs, especially in complex software engineering tasks, may ``hallucinate'', produce incorrect, incomplete, or misleading outputs~\cite{yang2024harnessing}. While LLMs are continuously improving, it is expected that they will eventually hit a ceiling. As discussed in our framework, automated checking and validation of LLM-generated outputs (possibly involving non-AI tools) will be essential to address these limitations. For instance, combining LLM code generation with a compiler to verify correctness is a promising trend that is already emerging~\cite{bi2024iterative}. 

\medskip
\noindent \textbf{Homogeneity of Code:} A potential issue in the future is that as more developers integrate AI into their workflows, LLMs may increasingly train on LLM-generated code. This could reduce the diversity of code practices over time, as LLMs tend to reaffirm patterns seen in their training data (e.g., public GitHub repositories). While this could make code more uniform and easier to understand for humans, it might also prevent exploration of alternative or potentially better coding practices. Additionally, such bias could lead to slow or insecure code being perpetuated by LLMs. 

\medskip
\noindent  \textbf{Ethical and Fairness Concerns:} Being trained on large amounts of data, LLMs are inherently biased toward common software practices and ways of thinking~\cite{huang2023bias}, which may not fully consider cultural or gender diversity. For example, previous research has found gender differences in various cognitive facets including information processing style with women more likely to prefer a comprehensive approach and have all information available before starting a task~\cite{burnett2016gendermag}. It is unknown if such cognitive diversity leads to differences in code style. This risk may be amplified by the above risk of homogeneous code, where the dominant way of thinking can be proliferated as code becomes more homogeneous. This is a general risk of AI, and addressing it is a critical challenge that requires effort from the AI community. Fine-tuning SE-specific LLM models to include sensitivity to ethical and fairness issues is an important step in mitigating this risk. More research is also needed to understand diversity of thought in relation to software design and code and how this can be maintained, or even broadened through increased participation, with LLMs. 
%\valerio{what I wrote is a bit vague, what else we could do?}\kelly{I added some things, could also reference some of the GenderMag work if you want an example of gender differences and use of software}\valerio{that would be good, thanks}\kelly{feel free to modify if it doesn't make sense to you}

\medskip
\noindent \textbf{Data Privacy:} LLMs services like OpenAI have the potential to memorize sensitive or proprietary information and  expose it to other users. Additionally, input data could be used as training data for future versions of the model~\cite{yao2024survey,yan2024protecting}. This is a serious risk, prompting many companies to use local or private LLM instances to ensure data privacy. We believe local LLM instances will become increasingly common, although they raise sustainability concerns. A centralized system (e.g., OpenAI models) is typically more cost-efficient for large-scale usage. However, private LLMs can be trained or fine-tuned with project- and company-specific data, enhancing performance for specialized tasks.

\subsection*{AI-General Limitations and Risks} 

\medskip
\noindent \textbf{Over-Reliance on AI:} Relying too heavily on AI may reduce developers’ problem-solving and critical thinking skills, as they become overly dependent on AI-generated solutions. This can lead to a loss of software engineering skills if humans become reliant on AI. This is of specific concern given the non-deterministic nature of AI. For example, reliance on a calculator, which was also controversial when first introduced, will always produce the same output given the same inputs, but the same is not true for AI systems. Software engineering skills are still needed to ensure quality of software systems. Similarly, people without software engineering expertise, may be overly confident in their abilities when leveraging LLMs. There is a broader risk that software developers might write less code over time, as their job could mainly become reviewing code generated by AI. This could lead to frustration and less sense of ownership over the codebase.

\medskip
\noindent \textbf{Malicious Activities:} Bad actors might exploit LLMs to generate malicious code or automate unethical practices, posing significant security risks. This is another reason why humans must remain in control and maintain their software engineering skills. 

\medskip
\noindent 
\textbf{Computational Requirements:} The high computational demands of AI (especially LLMs) can increase costs and environmental impact~\cite{samsi2023words}. However, recent advancements have shown that smaller, more efficient models can achieve good performance (e.g., see \textsc{GPT-4o-mini}~\footnote{\url{https://openai.com/index/gpt-4o-mini-advancing-cost-efficient-intelligence/}}). We believe the key lies in determining task complexity automatically and delegating only the most complex tasks to expensive LLM models~\cite{juneja2023small}.

%\valerio{anything else?}
%\kelly{maybe something also about license adherence? and more transparency - where did code recommendations come from? also maybe something on human aspects? AIs impact on developer experience, development becoming more like code review - which devs previously didn't like as much as writing code itself - how to keep motivation? this can also lead to loss of skill (because of turnover) if job satisfaction lowers. maybe something else on maintainability of code? what if code becomes harder to understand because humans aren't writing it - how to maintain it? how to ensure expertise? relates to the over reliance on AI item}

\section{Conclusions}

This paper presented a \textbf{vision of a symbiotic partnership between AI and software developers} motivated and inspired by recent advances in AI. This paper also discussed some key research challenges that need to be addressed by the software engineering community. While this paper focuses on specific software engineering challenges, it is essential to acknowledge broader AI-related concerns such as security, safety, bias, and privacy. Although not covered here, these issues are crucial but fall more within the domain of the AI community, and hopefully will be addressed soon.

\smallskip
We cannot ignore the opportunities that lie ahead. Nor should we disregard the concerns associated with them. Specifically, we must exercise caution \textbf{against over-reliance on AI}. While the next generations of software engineers should be trained in prompt engineering and AI, this should not overshadow the necessity of core software engineering knowledge. Human judgment remains indispensable for critically assessing AI-generated artifacts. It is crucial to emphasize again that AI serves as a tool to enhance developers' productivity and cannot (in the near future) replace humans. Putting too much trust on the software artifacts generated by AI can have serious repercussions on the quality and safety of our software systems~\cite{pearce2022asleep,yetistiren2023evaluating}.

\smallskip
 This paper serves also as a \textbf{call to arms for our community}. We need multi-disciplinary collaborations across our community to address the key challenges and achieve the envisioned symbiotic partnership between human developers and AI. While our vision is ambitious, we believe that a five to ten-year time frame is reasonable for realizing it.
 
\section*{Acknowledgments}
This work was supported by the Marsden Fund Council from Government funding, administered by the Royal Society Te Apārangi, New Zealand. 

\bibliographystyle{ACM-Reference-Format}
\bibliography{bibliography/references}

\end{document}